\documentclass[epj]{svjour}
\usepackage{graphics}
\usepackage{footnote}
\usepackage{tablefootnote}

\begin{document}

\title{A revised $B(E2; 2^+_1 \to 0^+_1)$ value in the semi-magic nucleus $^{210}$Po}
\titlerunning{A revised $B(E2; 2^+_1 \to 0^+_1)$ value in $^{210}$Po}

\author{D.~Kocheva\inst{1}, G.~Rainovski\inst{1}, J.~Jolie\inst{2}, N.~Pietralla\inst{3}, A.~Blazhev\inst{2}, A.~Astier\inst{4}, R.~Altenkirch\inst{2}, S.~Ansari\inst{2}, Th.~Braunroth\inst{2}, M.L.~Cort\'es\inst{3}, A.~Dewald\inst{2}, F.~Diel\inst{2}, M.~Djongolov\inst{1}, C.~Fransen\inst{2}, K.~Gladnishki\inst{1}, A.~Hennig\inst{2}, V.~Karayonchev\inst{2}, J.M.~Keatings\inst{5}, E.~Kluge\inst{2}, J.~Litzinger\inst{2}, C.~M\"uller-Gatermann\inst{2}, P.~Petkov\inst{6}, M.~Rudigier\inst{2}, M.~Scheck\inst{5}, P.~Spagnoletti\inst{5}, Ph.~Scholz\inst{2}, M.~Spieker\inst{2}, C.~Stahl\inst{3}, R.~Stegmann\inst{3}, M.~Stoyanova\inst{1}, P.~Th\"ole\inst{2}, N.~Warr\inst{2}, V.~Werner\inst{3}, W.~Witt\inst{3}, D.~W\"olk\inst{2}, K.O.~Zell\inst{2}, P.~Van Isacker\inst{7}, V.Yu.~Ponomarev\inst{3} %
}                     
\authorrunning{D. Kocheva {\it et al.}}

%
\institute{Faculty of Physics, St. Kliment Ohridski University of Sofia, 1164 Sofia, Bulgaria \and Institut f\"ur Kernphysik, Universit\"at zu K\"oln, 50937 Cologne, Germany \and Institut f\"ur Kernphysik, Technische Universit\"at Darmstadt, 64289 Darmstadt, Germany \and CSNSM, IN2P3/CNRS and Universit´e Paris-Sud, F-91405 Orsay Campus, France \and  University of the West of Scotland, PA1 2BE Paisley, UK and SUPA, Glasgow G12 8QQ, UK \and National Institute for Physics and Nuclear Engineering,77125 Bucharest-Magurele, Romania \and Grand Acc\'el\'erateur National d'Ions Lourds, CEA/DRF-CNRS/IN2P3, Bd. Henri Becquerel BP 55027, F-14076 Caen, France}
\date{Received: \today}
%
\abstract{
The lifetimes of the $2^+_1$, the $2^+_2$ and the $3^-_1$ states of $^{210}$Po have been measured in the $^{208}$Pb($^{12}$C,$^{10}$Be)$^{210}$Po transfer reaction by the Doppler-shift attenuation method. The results for the lifetime of the $2^+_1$ state is about three times shorter than the adopted value. However, the new value still does not allow for consistent description of the properties of the yrast $2^+_1$, $4^+_1$, $6^+_1$, and $8^+_1$ states of $^{210}$Po in the framework of nuclear shell models. The Quasi-particle Phonon Model (QPM) calculations also cannot overcome this problem thus indicating the existence of a peculiarity which is neglected in both theoretical approaches.  
} 
\maketitle
\section{Introduction}
\label{intro}
The low-energy spectra of semi-magic nuclei, especially those with two additional nucleons with respect to a doubly magic core, are of importance because they reveal the basic interactions that play a role in the formation of more complex collective excitations in open-shell nuclei. The properties of semi-magic nuclei are dominated by their single-particle structure and can be well understood in the framework of the nuclear shell model~\cite{Mayer}, which constitutes the standard approach to understand their low-energy spectra. Besides single-particle structure, pairing correlations are crucial in these nuclei because they favour the formation of configurations with nucleons in pairs coupled to angular momentum $J=0$. As a result semi-magic nuclei have low-energy excitations of low `seniority' $\upsilon$, the quantum number that counts the number of unpaired nucleons \cite{Racah43}. These nuclei display multiplets of states that have the same seniority $\upsilon$. If the valence nucleons are confined to a single-$j$ orbital, it is known that, to a good approximation, the interaction between identical nucleons conserves seniority \cite{Shalit63,Talmi93}. If the valence shell consists of several orbitals, a generalized-seniority scheme can be formulated \cite{Talmi71}, which can be considered as a truncation of the nuclear shell model.

The conservation of seniority imposes a number of simple rules \cite{Shalit63,Talmi93,Casten,Ressler04} that can be exploited for tracing the evolution from seniority-like behaviour to collectivity with increasing number of valence nucleons~\cite{Ressler04-2,Grahn}. Nuclei with two valence nucleons serve as reference points, providing the properties of the basic $j^2$ configurations such as the energies of states with $\upsilon=2$ and the absolute $E2$ transition strengths for the seniority-changing transition ($\Delta\upsilon=2$) $2^+_1\rightarrow0^+_1$ and the seniority-preserving transitions ($\Delta\upsilon=0$) within the $\upsilon=2$ multiplet. Moreover, these observables can be used for basic tests of complete shell-model calculations.

\begin{figure}
\resizebox{0.48\textwidth}{!}{%
  \includegraphics{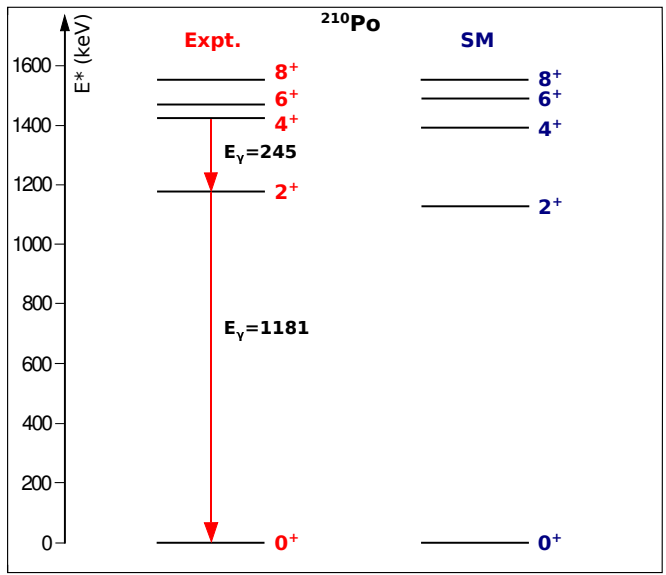}}
\caption{(Color online) Experimental and calculated yrast states of $^{210}$Po. The calculated shell-model levels are taken from Ref. \cite{Coraggio}.}
\label{exp-sm}       
\end{figure}

The semi-magic nucleus $^{210}$Po has two additional protons with respect to the doubly-magic nucleus $^{208}$Pb. The energies of its yrast $2^+$, $4^+$, $6^+$, and $8^+$ states follow a seniority-like pattern of decreasing energy splitting between adjacent states with increasing spin (cf. Fig.~\ref{exp-sm}), suggesting that the states belong to the $(\pi h_{9/2})^2$ multiplet. Data on absolute strengths for $E2$ transitions between these states are also available. The data are extracted from the lifetimes of the $4^+$, $6^+$ and $8^+$ levels, which are measured by the electronic timing technique \cite{4+state1}. The $B(E2;2^+_1\rightarrow0^+_1)$ value is determined from the cross-sections for populating the $2^+_1$ state in $^{210}$Po in inelastic scattering of deuterons and protons \cite{Ellegaard}. Large-scale shell-model studies of $^{210}$Po using realistic interactions \cite{Coraggio,Caurier} excellently reproduce the energies of the yrast $2^+$, $4^+$, $6^+$, and $8^+$ levels as can be seen in Fig.~\ref{exp-sm}. The calculated wave functions show that the dominant component of the yrast states is the $(\pi h_{9/2})^2$ configuration, corroborating the expectation that they belong to the seniority $\upsilon=2$ multiplet. The $E2$ strengths for the transitions between the $4^+$, $6^+$ and $8^+$ states are also almost perfectly reproduced (cf.\ Table VII in Ref. \cite{Coraggio} and Table III in Ref. \cite{Caurier}). However, in both studies \cite{Coraggio,Caurier} the $B(E2;2^+_1\rightarrow0^+_1)$ value is overestimated by a factor of six. Such a significant discrepancy between the shell-model calculations and the data is an indication for either an inaccurate experimental value \cite{Coraggio,Caurier} or for deficiencies in the model, as suggested in Ref. \cite{Caurier}. This has motivated us to perform a model-independent measurement of the lifetime of the $2^+_1$ level in $^{210}$Po based on the Doppler-shift attenuation method (DSAM)(cf.\ Ref. \cite{Alexander} and references therein) and to seek an alternative theoretical description of the properties of the yrast states of $^{210}$Po in the framework of the Quasi-particle Phonon Model (QPM) \cite{Soloviev}.    

\section{Experimental set-up}
\label{exp}
The experiment was performed at the FN Tandem facility of the University of Cologne. Excited states of $^{210}$Po were populated using the $^{208}$Pb($^{12}$C,$^{10}$Be)$^{210}$Po transfer reaction at a beam energy of 62 MeV. The beam energy was chosen
to be about 2 MeV below the Coulomb barrier. The target was a self-supporting 10 mg/cm$^{2}$ thick Pb foil enriched to 99.14 \% with the isotope $^{208}$Pb. The reaction took place in the reaction chamber of the Cologne plunger device \cite{Dewald}. In order to detect the recoiling light reaction fragments an array of solar cells was used. The array consisted of six 10 mm $\times$ 10 mm cells and was placed at a distance of about 15 mm between their centers and the target covering an angular range between 116.8$^{\circ}$  and 167.2$^{\circ}$. The emitted $\gamma$-rays were registered by 11 HPGe detectors mounted outside the plunger chamber in two rings at an average distance of 12 cm from the target. Five detectors were positioned at 142.3$^{\circ}$ with respect to the beam direction and the other six formed a ring at 35$^{\circ}$. Data were taken in coincidence mode of at least one solar cell and one HPGe detector (particle-$\gamma$) or when at least two HPGe detectors ($\gamma - \gamma$) have fired in coincidence.

\section{Data analysis and results}
\label{analysis}

\begin{figure*}
\resizebox{1.00\textwidth}{!}{%
  \includegraphics{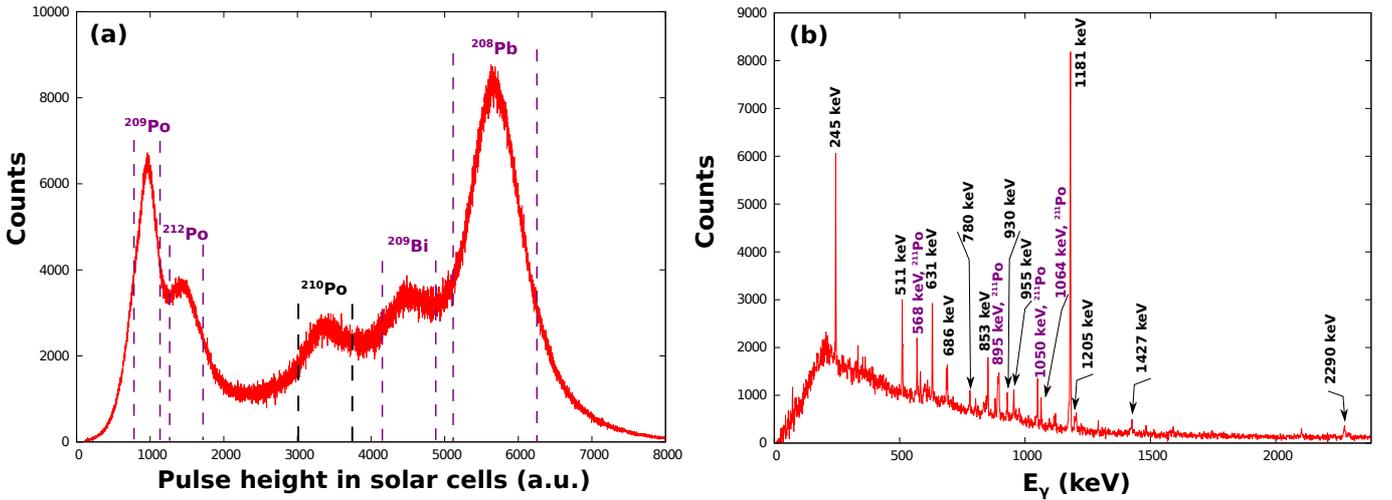}
}
\caption{(Color online) (a) The projection of the particle-$\gamma$ matrix obtained by coincidence detection of charged particles in the solar-cell array and a $\gamma$ ray at $\Theta _\gamma = 142^\circ$ polar angle. The vertical dashed lines represent parts of the particle spectrum found to be in coincidence with $\gamma$-rays from the indicated nuclei. (b) The $\gamma$-ray spectrum in coincidence with the group of particles indicated as "$^{210}$Po" in panel (a).}
\label{SC-gamma}       
\end{figure*}

The particle-$\gamma$ coincidence data were sorted in two matrices depending on the position of the HPGe detectors. A projection of the particle-$\gamma$ matrix obtained with $\gamma$-ray detection at $142^\circ$ is shown in Fig.~\ref{SC-gamma}(a). The $\gamma$ rays in coincidence with the group of particles indicated as "$^{210}$Po" in Fig.~\ref{SC-gamma}(a) are shown in Fig.~\ref{SC-gamma}(b). This spectrum is dominated by the 1181-keV and the 245-keV lines which are the $\gamma$-ray transitions depopulating the first two yrast states of $^{210}$Po~\cite{Mann} (cf. Fig.~\ref{exp-sm}). Besides some contaminants from $^{211}$Po (which are shown in purple), all other $\gamma$ rays in the spectrum in Fig.~\ref{SC-gamma}(b) originate from the decay of excited states of $^{210}$Po. The 1181.4-, 2290.1-, and 1205.4-keV $\gamma$-ray lines show well-pronounced Doppler shapes which allow us to extract the lifetimes of the $2^+_1$, the $2^+_2$ and the $3^-_1$ states of $^{210}$Po, respectively.   

The line-shape analysis was performed with the integrated software package APCAD (Analysis Program for Continuous Angle DSAM) \cite{Stahl}. In APCAD, the slowing down process is simulated by GEANT4 \cite{geant4}. The electronic stopping powers were taken from the Northcliffe and Schilling tables \cite{Northcliffe70} with corrections for the atomic structure of the medium, as discussed in Ref. \cite{Ziegler85}. The angular straggling due to nuclear collisions is modelled discretely by means of Monte Carlo simulation while the corresponding energy loss is considered to emerge as a result from a continuous process for which the nuclear stopping powers were taken from SRIM2013 \cite{Ziegler10} and reduced by 30\% \cite{Keinonen85}. The analysis accounts for the response of the HPGe detectors, for the experimental geometry, and for the restrictions on the reaction kinematics imposed by the solar-cell array. The software and the approach described above were verified in a previous study \cite{Kocheva}.

\begin{figure}
\resizebox{0.48\textwidth}{!}{%
  \includegraphics{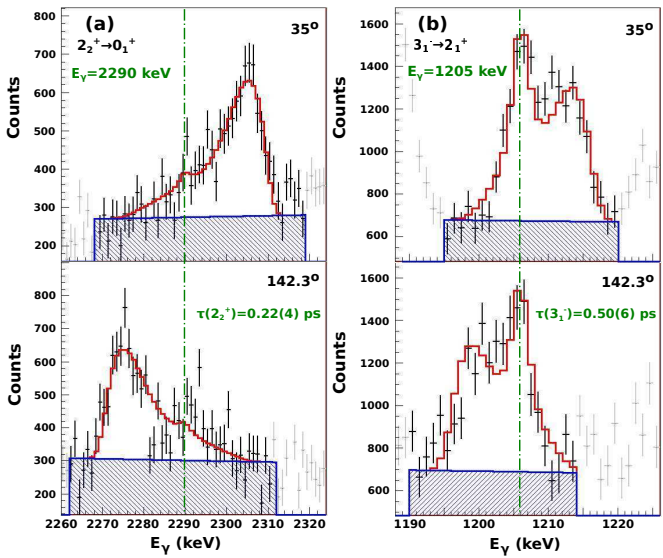}
}
\caption{(Color online) An example of line-shape fits of the 2290.1-keV ($2^+_2\rightarrow 0^+_1$) (a) and the 1205.4-keV ($3^-_1\rightarrow 2^+_1$) (b) transitions. The crossed (blue) areas show the background regions. The solid (red) lines represent the fit. The vertical dash-dotted (green) lines show the position of the unshifted peaks.}
\label{2290-1205}       
\end{figure}

The lifetimes of the $2^+_2$ state at 2290 keV and $3^-_1$ state at 2387 keV were extracted from the line shapes of the 2290-keV ($2^+_2\rightarrow 0^+_1$) and the 1205-keV ($3^-_1\rightarrow 2^+_1$) transitions in particle-gated spectra, respectively. These spectra are, in fact, $\gamma$-ray singles spectra which, in principal, only contain information for the effective lifetimes. However, due to the reaction mechanism it is justified to consider that slow feeding contributions to the effective lifetimes of excited states of $^{210}$Po can originate only from discrete decays of higher-lying states, as suggested in Ref. ~\cite{Astier10}. The $\gamma - \gamma$ coincidence data show that the  1205-keV transition is in coincidence with the 1181-keV ($2^+_1\rightarrow 0^+_1$) transition only, while the 2290-keV ($2^+_2 \rightarrow 0^+_1$) transition is not present in the coincidence data. This observation indicates that the $2^+_2$ and the $3^-_1$ states are directly populated in the transfer reaction. Therefore only fast feeding ($\tau _{feeding} \le 10$ fs) was introduced in the fits of their line shapes. The fits are presented in Fig.~\ref{2290-1205} and the extracted lifetimes are summarized in Table~\ref{final-values}.

\begin{figure}
\centering
\resizebox{0.48\textwidth}{!}{%
  \includegraphics{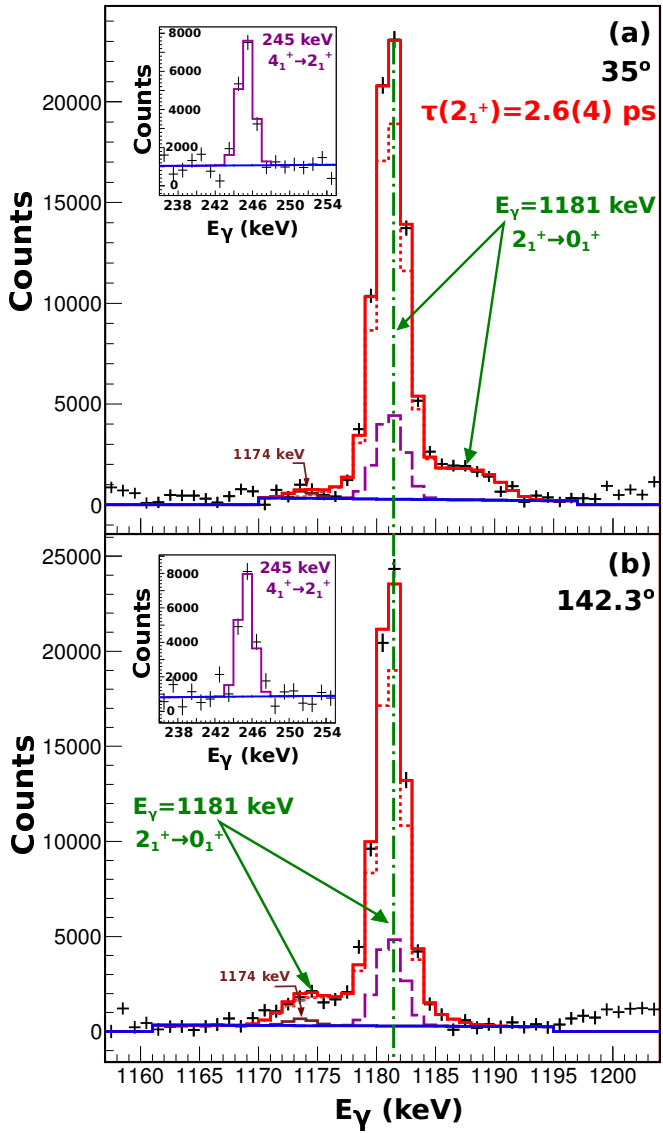}
}
\caption{(Color online) Simultaneous line-shape fits of the 1181-keV ($2^+_1\rightarrow 0^+_1$) transition observed at forward (a) and at backward (b) angles. The solid (red) line represents the total fit. The 1181-keV line is fitted simultaneously with the 245-keV ($4^+_1\rightarrow 2^+_1$) line which is always emitted from a stopped nucleus (the insets). The dotted and dashed lines represent the individual contributions of 1181-keV (red) and 245-keV (purple) lines, respectively, to the total fit. An unidentified stopped contaminant with $E_\gamma =1174$ keV is taken into account (brown).}
\label{1181}       
\end{figure}

The lifetime of the first excited $2^+$ state of $^{210}$Po was obtained from the line shape of the 1181-keV ($2^+_1\rightarrow 0^+_1$) transition. Using the $\gamma - \gamma$ coincidence data and the known branching ratios \cite{Mann} we have estimated that 38\% of the feeding of the $2^+_1$ state comes from the decay of the $4^+_1$ state at 1426 keV (20\%), the $2^+_2$ state at 2290 keV (2\%), and the $3^-_1$ state at 2387 keV (16\%). The lifetimes of the $2^+_2$ and the $3^-_1$ state (cf. Table~\ref{final-values}) and their influence of the line shape of the 1181-keV transition were taken into account by the fitting procedure for the lifetime of the $2^+_1$ state. A special care was taken to account for the impact of the 245-keV ($4^+_1\rightarrow 2^+_1$) transition. The $4^+_1$ state of $^{210}$Po is a long-lived state with lifetime $\tau =1.56(6)$ ns \cite{4+state1,4+state2} and in the present experiment it always decays at rest. Hence, when the $2^+_1$ state is fed from the $4^+_1\rightarrow 2^+_1$ transition, it also always decays at rest which gives extra counts into the fully stopped component of the 1181-keV transition. In order to extract correctly the lifetime of the first $2^+$ state of $^{210}$Po by means of the Doppler-shift attenuation method, the contribution of the $\gamma$ rays coming from the 245-keV transition to the fully stopped component of the 1181-keV transition has to be eliminated. The efficiency corrected number of counts in the 245-keV line are subtracted from the efficiency corrected number of counts in the stopped component of the 1181-keV transition. That procedure could be automatically carried out with APCAD by simultaneously fitting the 1181- and the 245-keV lines. In addition, the $\gamma - \gamma$ coincidence data indicates the presence of 1427-keV transition which connects the $0^+_2$ state at 2608 keV to the $2^+_1$ state. However, this is an extremely weak transition which accounts for less than 1\% of the total population of the $2^+_1$ state and practically, has no impact on the lifetime of the $2^+_1$ state. Therefore, besides the influence of the feeding from the $4^+_1$, the $2^+_2$ and the $3^-_1$ states, only fast feeding ($\tau_{feeding} \le 10$ fs) was considered in the fit of the line shape of 1181-keV transition. Under this assumption, the final value of the lifetime of the $2^+_1$ is extracted from a simultaneous line-shape fit of the 1181-keV transition observed at forward and backward angles (see Fig.~\ref{1181}) and presented in Table~\ref{final-values}.

\begin{table*}
\begin{minipage}{\textwidth}
\centering
\caption{Properties of the three investigated low-lying states of $^{210}$Po and $\gamma$-ray transitions originating from their decays. Denoted are: the excitation energies ($E_{level}$); the spin and parity of the initial levels ($J^\pi$) and of the final levels ($J^\pi_{final}$); the $\gamma$-ray energies ($E_\gamma$); the relative intensities ($I_\gamma$); the total electron conversion coefficients ($\alpha$); the multipole mixing ratios ($\delta$) of the $\gamma$-ray transitions; the lifetimes of the states; the absolute transition strengths.}
\label{final-values}      
\begin{tabular}{lllllllll}
\hline\noalign{\smallskip}
$E_{level}$ & $J^\pi$ & $J^\pi_{final}$ & $E_\gamma$ & $I_\gamma$ \footnote{From Ref.~\cite{Mann}.} & $\alpha$ \footnote{Total electron conversion coefficients. From Ref.~\cite{nndc}.} & $\delta$ \footnote{From Ref.~\cite{Mann}.} & $\tau$ (ps) & Transition strength \\
(keV) & & & (keV) & \% & & & & $J^\pi \to J^\pi_{final}$ \footnote{$B(E1)$ values are given in $e^2$fm$^2$, $B(E2)$ values are given in $e^2$fm$^4$ (1 W.u.=74.15 $e^2$fm$^4$), $B(E3)$ value is given in $e^2$fm$^6$, and $B(M1)$ value is given in $\mu_N^2$.} \\
\noalign{\smallskip}\hline\noalign{\smallskip}
1181 & $2^+_{1}$ & $0^+_{1}$ & 1181 & 100 & 0.00535 & & 2.6(4) & $B(E2)$=136(21)\\
\noalign{\smallskip}\hline\noalign{\smallskip}
2290 & $2^+_{2}$ & $0^+_{1}$ & 2290 & 100(2) & 0.00198 & & 0.22(4) & $B(E2)$=53(11)\\
& & $2^+_{1}$ & 1108 & 11.2(11) &  0.0133 & 0.61(31) & & $B(M1)$=0.014(7)\\
& & & & & & & & $B(E2)$=60(29)\\
\noalign{\smallskip}\hline\noalign{\smallskip}
2387 & $3^-_{1}$ & $0^+_{1}$ & 2387 & 1.0(3) & 0.00309 & & 0.50(6) &  $B(E3)$=70(31)$\times$10$^3$\\
& & $2^+_{1}$ & 1205 & 100.0(15) & 0.00197 & & & $B(E1)$=0.64(9)$\times$10$^{-3}$ \\
& & $4^+_{1}$ & 960 & 11.3(6) & 0.00292 & & & $B(E1)$=0.14(3)$\times$10$^{-3}$ \\
\noalign{\smallskip}\hline
\end{tabular}
\end{minipage}
\end{table*}

\section{Discussion}
\label{discussion}
The lifetimes from the present study, together with the available spectroscopic information and the resulting transition strengths, are summarized in Table~\ref{final-values}. The lifetime for the $3^-_1$ state of $^{210}$Po is in a good agreement with the estimated value from Ref. \cite{Ellegaard}. The lifetime of the $2^+_2$ state is measured for the first time. The new value for the $2^+_1$ state is about a factor of three shorter than the adopted one \cite{nndc}. The latter is deduced from a relative cross-section measurement in a $(d,d')$ scattering experiment \cite{Ellegaard}. It has to be noted, however, that the new value is in a better agreement with the lifetime resulting from the cross-section for $(p,p')$ scattering reported in the same study \cite{Ellegaard}. On the other hand, the value from the present study is obtained by a model independent technique and free of the systematic uncertainties inherent to cross-section analysis. 

The shorter lifetime of the $2^+_1$ state leads to $B(E2;2^+_1 \to 0^+_1) = 1.83(28)$ W.u. which is about three times larger than the adopted value \cite{Ellegaard,nndc}. However, this is not sufficient to compensate for the discrepancy between the experimental value and the estimations within the pure seniority scheme \cite{Stuchbery} or the shell model result (cf. Table VII in Ref. \cite{Coraggio} and Table III in Ref. \cite{Caurier}) which remain two times larger than the experimental value. Since, at this point the discrepancy cannot be explained any longer as due to experimental uncertainties, it seems reasonable, to be attributed to the neglect of $ph$ excitations of the $^{208}$Pb core which strongly influence the $I^\pi = 2^+$ states \cite{Nowacki} as suggested in Ref. \cite{Caurier}.  

It is also interesting to check whether the problem is specific for shell models and whether other theoretical approaches can do better in describing the properties of low-lying states of $^{210}$Po. For this purpose we have performed Quasi-particle Phonon Model (QPM) calculations~\cite{Soloviev} for $^{210}$Po. In these calculations we have used the Woods-Saxon potential for the  mean field with the parameters presented in Ref.~\cite{BerPon}. The single-particle space includes all shells from the potential bottom to quasibound states with narrow widths. The single-particle energies of the levels near the Fermi surface are taken from Ref.~\cite{KhoVor}, where they have been adjusted to reproduce experimental energies of low-lying levels of four odd-mass neighbouring nuclei to $^{208}$Pb in a calculation with the interactions between quasiparticles and phonons accounted for. The strength of the monopole pairing interaction for protons in $^{210}$Po has been adjusted to the corresponding even/odd mass differences. Because of the large single-particle model space no effective charges were used for calculating the $E1$, $E2$, and $E3$ transition strengths. The spin-gyromagnetic quenching factor used to calculate the $M1$ transition strengths is $g_s^{\rm eff} = 0.8 g_s^{\rm free}$. Excitation modes of even-even nuclei (both collective and almost two-quasi-particle ones) are treated in the QPM in terms of phonons. Their excitation energies and internal fermion structure is obtained from solving Quasi-particle Random-Phase Approximation (QRPA) equations. The model uses a separable form of the residual interaction. In the present studies we have used Bohr-Mottelson form factor of the residual force as a derivative of the mean field potential. The strength of the isoscalar residual interaction, the same for all multipoles, has been adjusted to the experimental $B(E2, 2^+_1 \to 0_{g.s.}$ ) value in $^{210}$Po. The calculations have been performed on the basis of interacting one-~, two-, and three-phonon configurations. The phonons of the $2^+$, $3^-$, $4^+$, $6^+$, and $8^+$ multipolarities have been involved. Complex (two- and three-phonon) configurations have been built up from all possible combinations of these phonons. The basis has been truncated above 6~MeV, 8~MeV, and 10~MeV for one-, two-, and three-phonon components, respectively.

\begin{table*}
\begin{minipage}{\textwidth}
\centering
\caption{Results from the QPM calculations for the low-lying states of $^{210}$Po in comparison with the experimental data.}
\label{theory}      
\begin{tabular}{cccccll}
\hline\noalign{\smallskip}
\multicolumn{1}{c}{$J^\pi_i$} & \multicolumn{2}{c}{$E_x$, (MeV)} & Structure \footnote{[$2_1^+$]$_{RPA}$ means the lowest in energy RPA phonon of the multipolarity $2^+$, etc.} & \multicolumn{3}{c}{Transition strength \footnote{$B(E1)$ values are given in $e^2$fm$^2$, $B(E2)$ values are given in $e^2$fm$^4$ (1 W.u.=74.15 $e^2$fm$^4$),$B(E3)$ value is given in $e^2$fm$^6$, and $B(M1)$ value is given in $\mu_N^2$.}} \\
\cline{2-3} \cline{5-7}
 & Experiment & QPM & \% &$J^\pi_f$& Experiment & QPM \\
\noalign{\smallskip}\hline\noalign{\smallskip}
$2_1^+$ & 1.18 & 1.10 & 97\%[$2_1^+$]$_{RPA}$ & $0^+_1$ &$B(E2)$=136(21) &$B(E2)$=135 \\
\noalign{\smallskip}\hline\noalign{\smallskip}
$4_1^+$ & 1.43 & 1.16 & 99\%[$4^+_1$]$_{RPA}$ & $2^+_1$ &$B(E2)$=335(14) \footnote{From Ref.~\cite{4+state1,4+state2}.} & $B(E2)$=41 \\
\noalign{\smallskip}\hline\noalign{\smallskip}
$6_1^+$ & 1.47 & 1.20 & 99\%[$6^+_1$]$_{RPA}$ & $4^+_1$ &$B(E2)$=229(7) \footnote{From Ref.~\cite{4+state1}.} & $B(E2)$=28 \\
\noalign{\smallskip}\hline\noalign{\smallskip}
$8_1^+$ & 1.56 & 1.21 & 99\%[$8^+_1$]$_{RPA}$ & $6^+_1$ &$B(E2)$=84(11) \footnote{From Ref.~\cite{4+state1,Mann}.} & $B(E2)$=11 \\
\noalign{\smallskip}\hline\noalign{\smallskip}
$2_2^+$ & 2.29 & 2.05 & 1.2\%[$2^+_1$]$_{RPA}$ + 60.5\%[$2^+_1 \otimes 2^+_1$]$_{RPA}$ +  & $0^+_1$ &$B(E2)$=53(11) & $B(E2)$=3.4 \\
		&	   &		  &	 + 20.7\%[$2^+_1 \otimes 4^+_1$]$_{RPA}$ & $2^+_1$ &$B(M1)$=0.014(7) & $B(M1)$=0.006 \\
		&	   &		  &	& $2^+_1$ &$B(E2)$=60(29) & $B(E2)$=80 \\
\noalign{\smallskip}\hline\noalign{\smallskip}
$3^-$ & 2.39 & 2.61 & 95\%[$3^-_1$]$_{RPA}$ & $0^+_1$ &$B(E3)$=70(31)$\times$10$^3$ & $B(E3)$=70$\times$10$^3$ \\
		&	   &		& & $2^+_1$ &$B(E1)$=0.64(9)$\times$10$^{-3}$ & $B(E1)$=0.83$\times$10$^{-3}$ \\
		&	   &		& & $4^+_1$ &$B(E1)$=0.14(3)$\times$10$^{-3}$ & $B(E1)$=0.27$\times$10$^{-3}$ \\
\noalign{\smallskip}\hline
\end{tabular}
\end{minipage}
\end{table*}

The results from the calculations are presented and compared to the experimental data in Table~\ref{theory}. The energies of the states of interest are reasonably well reproduced. It has to be noted that in the chosen approach to fix the strength parameters to the electric strengths of the $2^+_1$  state, the result for the energies of the states should be considered as a prediction of the model. In this respect, the agreement between the experimental and the calculated energies for the $2^+_1$ and the $3^-_1$ states allows to interpret them as one-quadrupole and one-octupole phonon states, respectively. The energies of the $4^+_1$, $6^+_1$, and $8^+_1$ states are somewhat lower in the calculation than experimentally observed. It should be noted that the energy of the lowest two-quasiparticle configuration $\pi\{(1h_{9/2})^2\}$ is 1.211~MeV with the present quasiparticle spectrum. In general, the calculation predicts a rather pure one-phonon nature of the lowest $2^+$, $4^+$, $6^+$ and $8^+$ states.

The major discrepancy between the QPM calculations and the experimental data appears in the $E2$ transition strengths for the cascade $8^+_1 \to 6^+_1 \to 4^+_1$ (cf. Table~\ref{theory}). Overall, the model underestimates these values by a factor of 8. Since the corresponding states have a rather pure one-phonon nature, these transitions are determined by the $E2$ matrix elements $\langle [6_1^+]_{RPA} ||E2 || [8_1^+]_{RPA} \rangle$, etc. These matrix elements are much smaller as compared to the decays from two-phonon components $\langle [6_1^+]_{RPA} ||E2 || [2^+_1 \otimes 6^+_1]_{RPA} \rangle$, etc with an exchange of the $[2_1^+]_{RPA}$ phonon~\cite{Bosfob}. Experimental data definitely indicate strong admixture of two phonon components in the structure of the $4^+_1$, $6^+_1$, and $8^+_1$ states which is not reproduced in the QPM calculations. The problem existing in the shell model description also appears in a different form in the present QPM calculations. This may indicate that the $^{208}$Pb core is soft as this softness strongly enhances the $1ph$ excitations, or facilitates the mixing between the $2qp$ configurations. Obviously, neither shell models nor QPM account correctly for this peculiarity.              

\section{Conclusions}
\label{conclusions}\
In the present study the lifetimes of the $2^+_1$, the $2^+_2$ and the $3^-_1$ states of $^{210}$Po have been measured by the Doppler shift attenuation method. The newly established value of the lifetime of the $2^+_1$ state leads to an $E2$ transition strength which is about three times larger than the adopted value \cite{nndc}. However, the new value is still not high enough to allow for a consistent microscopic description of the properties of the yrast $2^+_1$, $4^+_1$, $6^+_1$, and $8^+_1$ states of $^{210}$Po using $^{208}$Pb as an inert core. Such description is well expected since $^{210}$Po comprises only two additional protons with respect to the doubly magic nucleus $^{208}$Pb. The problem previously known from shell model studies also appears in the performed QPM calculations. This observation and the new experimental value for the lifetime of the $2^+_1$ state prompt for a more thorough theoretical investigation of this problem.     

\begin{acknowledgement}
{\bf Acknowledgements:} D.K. acknowledges the support from Deutscher Akademischer Austauschdienst. This work was supported by the partnership agreement between the University of Cologne and University of Sofia, by the BgNSF under grant DN08/23/2016, by STFC, by the German-Bulgarian exchange program under grants PPP57082997 \& DNTS/01/05/2014, by the DFG under grant SFB 1245, and by the BMBF under grants 05P12RDCIB, 05P15RDFN1,9 and 05P15RDCIA.
\end{acknowledgement}

\end{document}